\documentclass[aps,prb,showpacs,twocolumn]{revtex4-1}
\usepackage{color,graphicx}
\usepackage{amsmath}
\usepackage{bm}
\usepackage{gensymb}
\usepackage{textcase}

\begin{document}

\title{
Penetration of a magnetic wall into thin ferromagnetic electrodes
of a nano-contact spin valve}

\author{M. Sasaki$^{1}$, S. Tanaka$^{1}$, Y. Norizuki$^{1}$, K. Matsushita$^{2}$,
J. Sato$^{3}$, and H. Imamura$^{4}$}

\affiliation{
$^{1}$Department of Applied Physics, Tohoku University, Sendai, 980-8579, Japan\\
$^{2}$Cybermedia Center (CMC), Osaka University, Osaka, 560-0047, Japan\\
$^{3}$Department of Physics, Ochanomizu University, Tokyo, 112-8610,Japan\\
$^{4}$Spintronics Research Center, AIST, Tsukuba 305-8568, Japan}

\date{\today}

\begin{abstract}
 We theoretically analyzed a magnetic wall confined in a nano-contact
 spin valve paying special attention to the penetration of the
 magnetic wall into thin ferromagnetic electrodes. 
 We showed that, compared with the Bloch wall, the penetration of the
 N{\'e}el wall is suppressed by increases of the demagnetization energy.
 We found the optimal conditions of the radius and height of the 
 nano-contact to maximize the power of the current-induced oscillation of the
 magnetic wall. We also found that the thermal stability of the Bloch
 wall increases when the nano-contact's radius increases or height decreases.
\end{abstract}
\maketitle

\begin{figure}[t]
\includegraphics[width=\columnwidth]{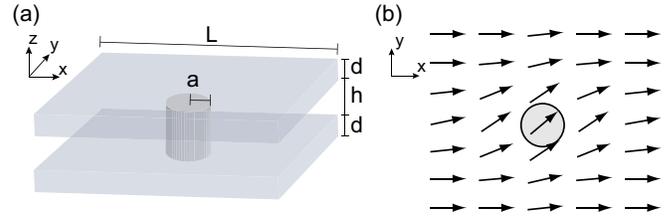}
\caption{(a) A nano-contact spin valve is schematically shown.
(b) A schematic illustration of the penetration of the magnetic wall into
 the ferromagnetic electrode. The shaded circle represents the area of
 the nano-contact.
} 
\label{fig:fig1}
\end{figure}

Magnetic walls have attracted much attention as a basic element of
nano-spintronics devices such as a racetrack memory\cite{hayashi2008},
spin-wave logic gates\cite{allwood2005}, a read-head for
ultra-high-density magnetic recording\cite{garcia1999,imamura2000,takagishi2009}, 
and a spin-torque oscillator 
(STO)\cite{He2007,Ono2008,franchin2008,suzuki2009,matsushita09B,Bisig2009}.
The nano-contact spin valve with a geometrically confined magnetic wall
shown in Fig.~\ref{fig:fig1}~(a) is a nanostructure suitable for the
latter two applications, and much effort has been devoted to studying its
electric and magnetic properties\cite{garcia1999,ono1999,bruno1999,Gorkom:1999,imamura2000,coey2001,chopra2005,suzuki2009}.
It is widely accepted that a magnetic wall should be created in the
nano-contact when the magnetization vectors of the top and bottom
electrodes are aligned to be anti-parallel. The magnetic wall in the
nano-contact can be categorized into two types: the Bloch wall and the
N{\'e}el wall. The magnetic moments of the Bloch (N{\'e}el) wall rotate
along the axis parallel (perpendicular) to the nano-contact\cite{matsushita09A,matsushita10,arai2012}.
The read head utilizes the resistance change due to the creation of a
magnetic wall, and the STO utilizes the oscillation between the Bloch wall and the
N{\'e}el wall induced by the applied direct current.

In early experiments\cite{garcia1999,ono1999,chopra2005} the break junction
of a ferromagnetic wire was
employed to realize the nano-contact spin valve, and the
corresponding theoretical analysis predicted that the magnetic walls were
almost completely confined in the contact region. The point is that the
energy required to rotate the magnetic moments in the contact is much
less than that in the electrode of the ferromagnetic wire.
However, if the thickness of the electrode is as small as the
height of the contact, the energy required to rotate the magnetic
moments in the electrode becomes so small that the magnetic wall can penetrate
into the electrode~\cite{Molyneux02,Jubert05,KohnSlastikov06} 
as shown in Fig \ref{fig:fig1}~(b). Recently the 
nano-contact spin valve with thin ferromagnetic electrodes~\cite{Fuke07} has been 
considered a powerful candidate for 2-5 Tb/in$^{2}$ 
read sensors\cite{takagishi2010}.
Therefore, it is important to analyze the penetration of the magnetic wall
into the thin ferromagnetic electrodes of the nano-contact spin valve.

In this paper, we derived analytical expression of a magnetic wall
penetrating into the thin ferromagnetic electrodes of a nano-contact
spin valve. Based on the derived analytical formula we studied the 
power of the current-induced oscillation and the thermal stability of
the magnetic wall.

The nano-contact spin valve we consider is schematically shown in
Fig. \ref{fig:fig1}~(a). To model the anti-parallel configuration of the
spin valve, we applied fictitious magnetic fields of magnitude $H$
to the top (bottom) electrode in the positive (negative) $x$-direction. 
The top and bottom electrodes were assumed to be square plates with width
$L$, depth $L$ and thickness $d$. The nano-contact was assumed to be a
cylinder with radius $a$ and height $h$. The electrodes and contact were
assumed to be made of the same ferromagnetic material. The $z$-axis was
taken to be parallel to the nano-contact and the electrodes parallel
to the $xy$-plane. The origin of the coordinates was set to be the center
of the nano-contact.

The direction of the magnetization can be expressed by the polar angle
$\theta$ and the azimuthal angle $\phi$.  
The energy in the contact is given by
\begin{equation}
\label{eq:energy_contact}
E_{\rm C}
 =\pi a^{2}\int_{-h/2}^{h/2} dz\, A
 \left[
 \left(
  \frac{\partial \theta}{\partial z}
 \right)^{2}
 +
 \sin^{2}\theta
 \left(
  \frac{\partial \phi}{\partial z}
 \right)^{2}
\right],
\end{equation}
where $A$ is the exchange stiffness constant. 
Because 
the exchange interaction energy is dominant in the nano-contact\cite{bruno1999}, 
we neglected the demagnetization and Zeeman energies.

The energy of the top electrode is given by
\begin{align}
E_{\rm T}= & 2\pi d\int_{a}^{\infty} dr\, r
 \left[
  A\left(
    \dot{\theta}^{2} +\sin^{2}\theta\,\dot{\phi}^{2}
   \right)
  \right.
 \nonumber\\
 & +
 \left.
  \frac{\mu_0 M_{\rm s}^2}{2}\cos^{2}\theta
  - \mu_0 M_{\rm s} H\left(\sin\theta\,\cos\phi -1 \right)
 \right],
\label{eq:energy_top_electrode}
\end{align}
where $r=\sqrt{x^{2} + y^{2}}$, $\dot{\phi}=\partial \phi / \partial r$,
$\dot{\theta}=\partial \theta / \partial r$, $M_{\rm s}$ is the saturation
magnetization, and $\mu_{0}$ is the permeability of the vacuum.
The first, second and third terms of Eq.~\eqref{eq:energy_top_electrode}
represent the exchange interaction energy, demagnetization energy, and
Zeeman energy, respectively.
The energy of the bottom electrode $E_{\rm B}$ is the same as $E_{\rm T}$.
Assuming that the $L$ is much larger than the penetration length of the
magnetic wall, the system can be regarded to be rotationally symmetric
about the $z$-axis; i. e., the angles $\phi$ and $\theta$ depend
only on the radius $r$. 

It is worth pointing out that, in the ground state, 
magnetization in the top electrode 
$\mbox{\boldmath $m$}_{\rm T}(r)$ and that in the bottom electrode $\mbox{\boldmath $m$}_{\rm B}(r)$ 
are related as $(-m_{\rm T}^x,m_{\rm T}^y,m_{\rm T}^z)=(m_{\rm B}^x,m_{\rm B}^y,m_{\rm B}^z)$ 
due to the symmetry of the system. In the following, we calculate the magnetic structure 
by taking this fact into account.


Let us first consider the Bloch wall, where we assume that
$\theta = \pi/2$, $d\theta/dz=0$, and $\dot{\theta}=0$. 
The energy of the contact is given by
\begin{equation}
\label{eq:ec_bloch}
E_{\rm C}
 =\pi a^{2}\int_{-h/2}^{h/2} dz\, A
 \left(\frac{\partial \phi}{\partial z}\right)^{2}
 =\frac{2\pi a^{2} A}{h}(\pi-2\phi_0)^{2}.
\end{equation}
To derive the last equation, we have assumed that $\phi|_{z=h/2}=\phi_0$ 
and $\phi|_{z=-h/2}=\pi-\phi_0$. We hereafter call $\phi_0$ the {\it boundary angle}. 
We have also used the fact that 
$\phi$ is a linear function of $z$. Note that the corresponding Euler equation is 
$\partial^2 \phi/\partial z^2=0$.
The energy of the top electrode is
\begin{equation}
E_{\rm T}
 =2\pi d\int_{a}^{\infty} dr\, r
 \left[
  A\dot{\phi}^{2} - \mu_0M_{\rm s} H\left(\cos\phi -1 \right)
 \right].
\label{eq:et_bloch}
\end{equation}
The corresponding Euler equation was obtained as
\begin{equation}
\label{eq:euler1}
\ddot{\phi}
 +
 \frac{1}{r}\dot{\phi}
 -
\frac{\mu_0 M_{\rm s} H }{2 A}\sin\phi
=0. 
\end{equation}
The angle $\phi$ at $r=a$ should be consistent with the boundary 
condition $\phi_{z=h/2}=\phi_0$ for the contact and magnetic moments align in 
the direction of the magnetic field at $r=\infty$. 
We therefore obtained the boundary conditions $\phi|_{r=a} = \phi_0$ and $\phi|_{r=\infty}=0$.
In the bottom electrode the boundary conditions
should be $\phi|_{r=a} = \pi-\phi_0$ and $\phi(\infty)=\pi$. 
The energy of the electrode, which is a function of $\phi_0$, 
was obtained by substituting the solution of Eq.~(\ref{eq:euler1}) 
into Eq.~(\ref{eq:et_bloch}) and performing the integral.
The boundary angle $\phi_0$ was obtained by minimizing the total energy 
$E_{\rm total} = E_{\rm C} + E_{\rm T}+E_{\rm B}$.

\begin{figure}[t]
    \begin{tabular}{cc}
      \resizebox{42mm}{32mm}{\includegraphics{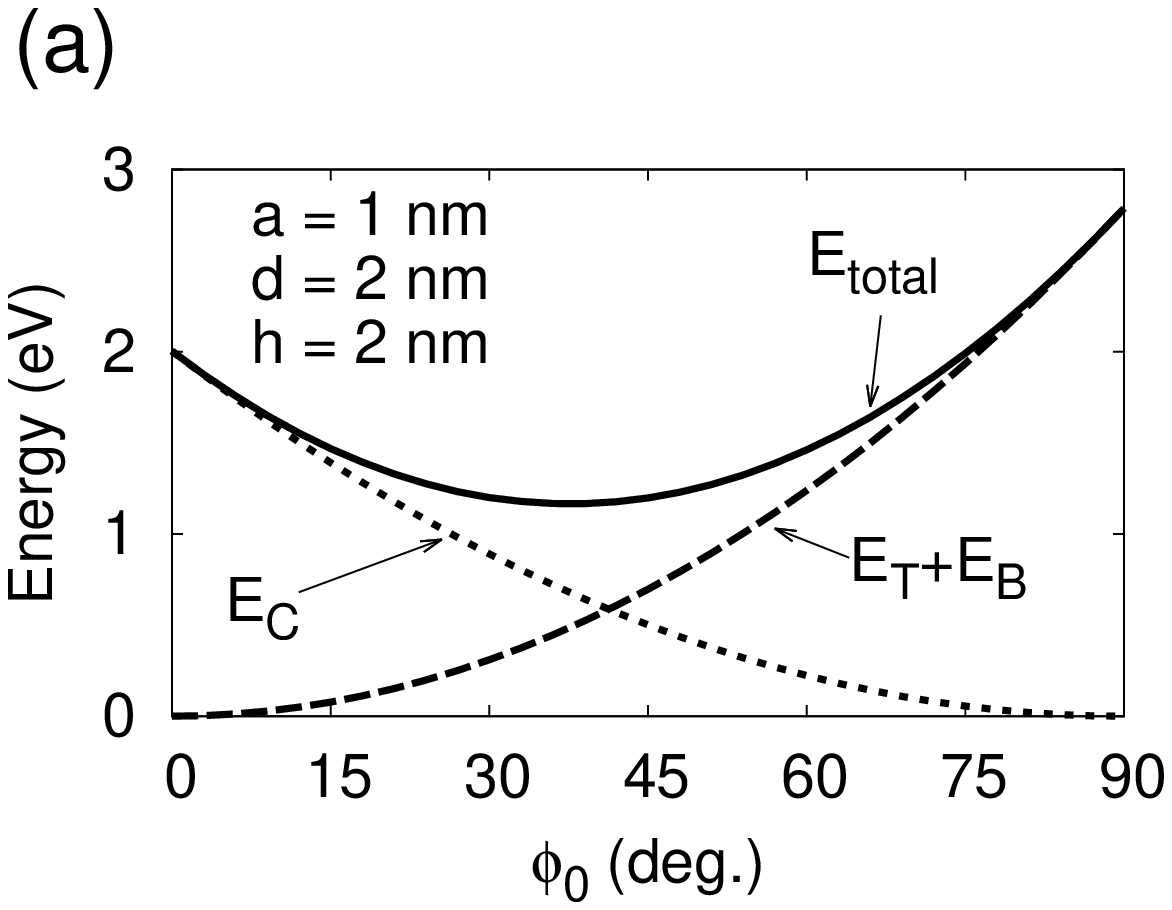}} &
      \resizebox{42mm}{32mm}{\includegraphics{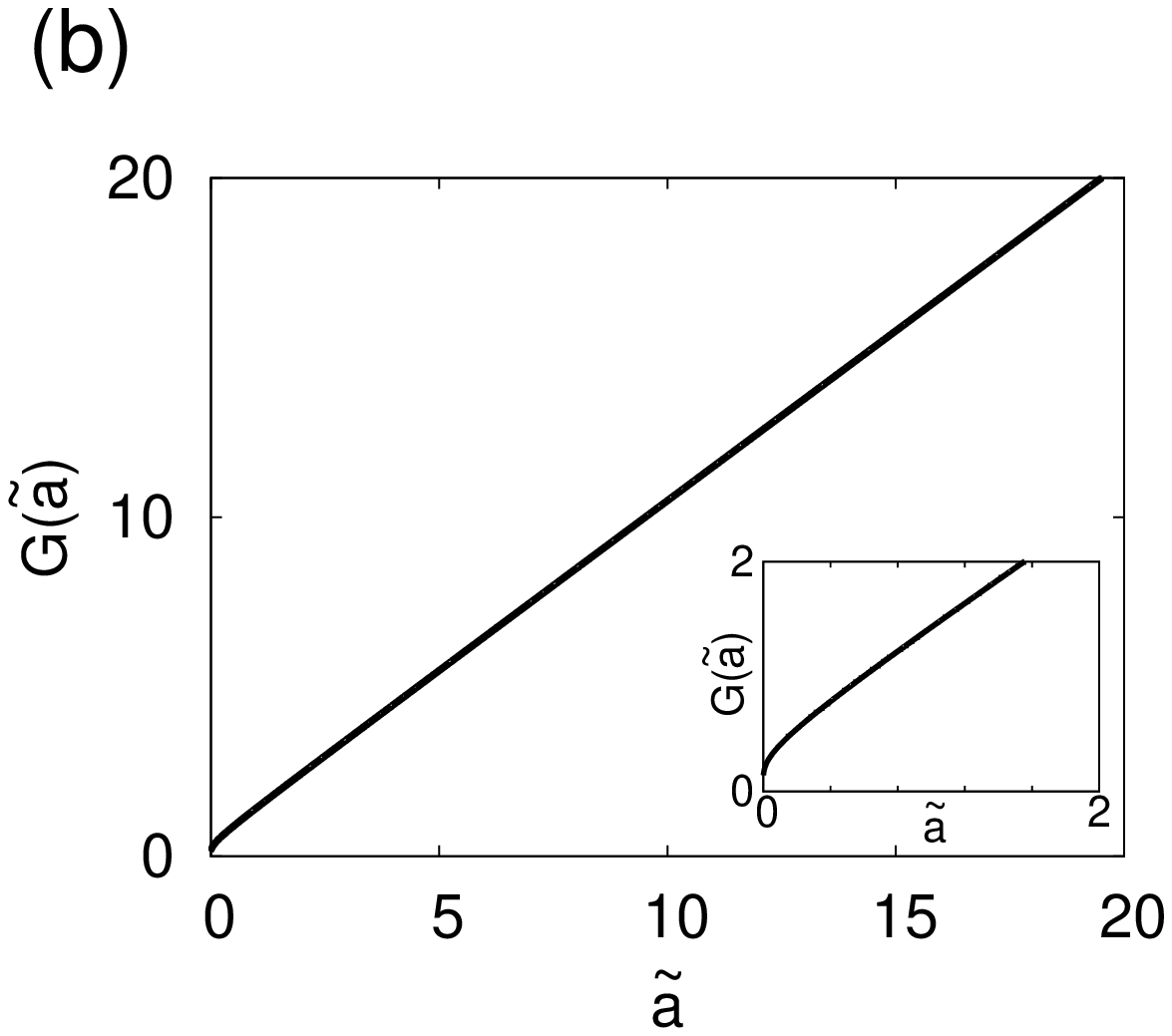}} \\
      \resizebox{42mm}{32mm}{\includegraphics{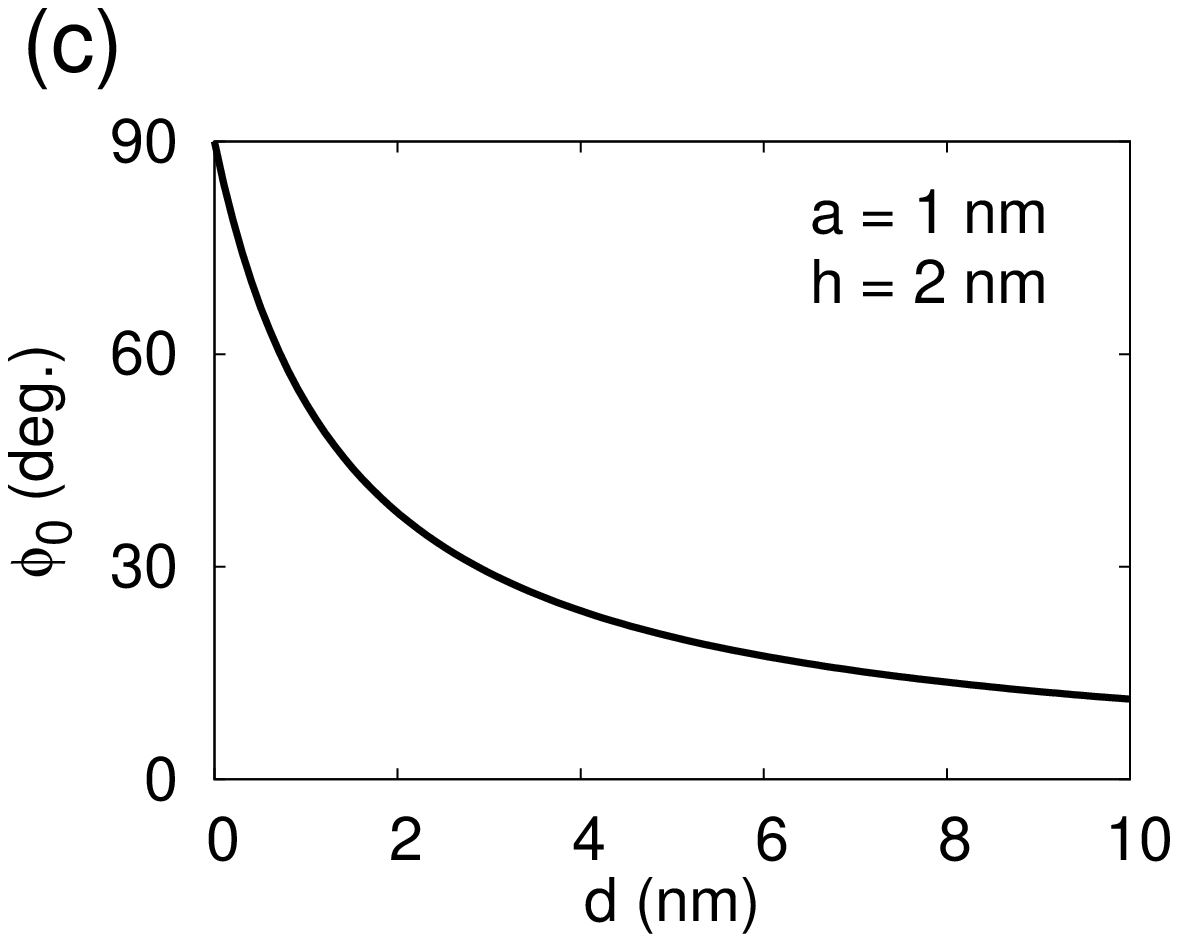}} &
      \resizebox{42mm}{32mm}{\includegraphics{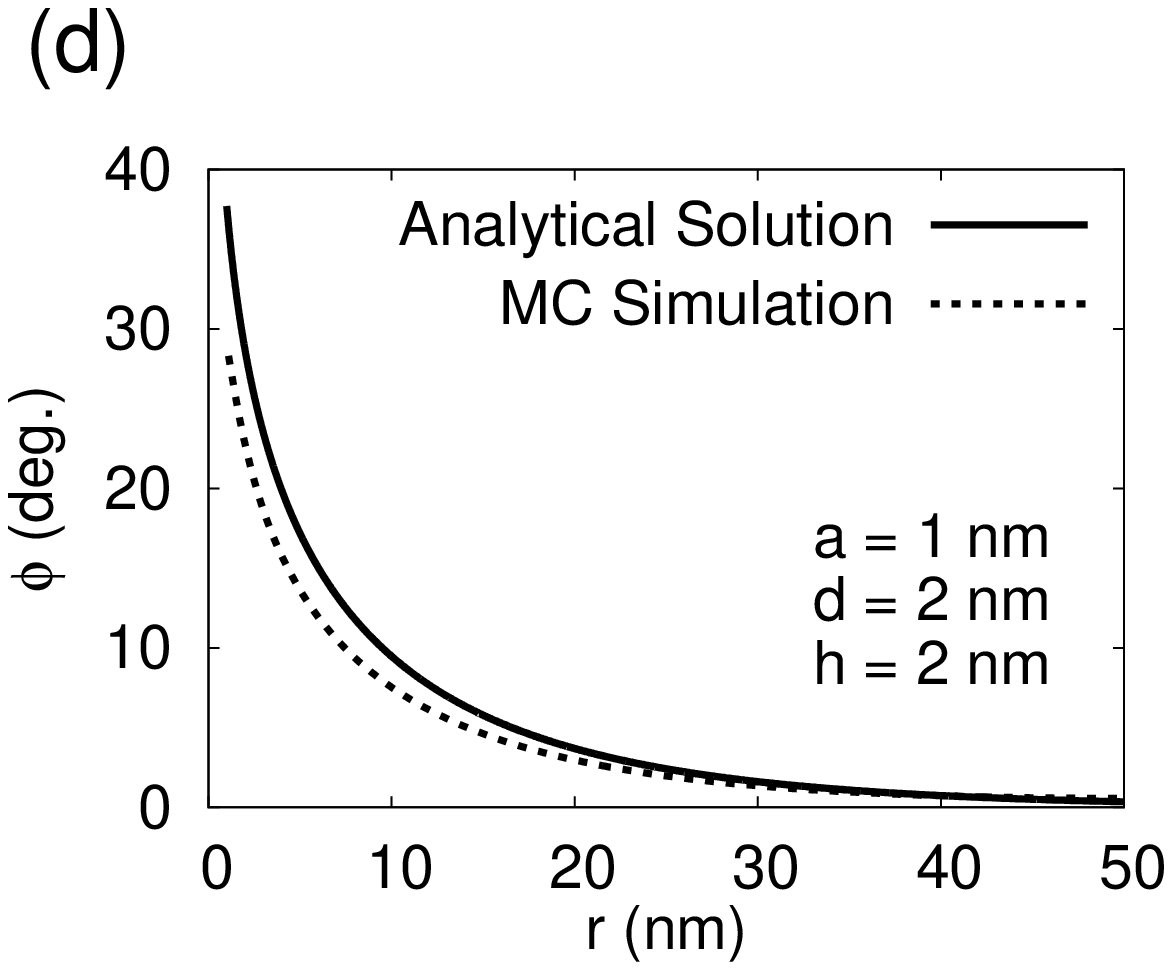}} \\
    \end{tabular}
\caption{
 (a) The energies of the contact $E_{\rm c}$, the top and bottom
 electrodes $E_{\rm T} + E_{\rm B}$, and the total energy 
 are plotted against the boundary angle $\phi_0$ by the dotted,
 dashed, and solid lines, respectively.
 (b) The function $G(\tilde{a})$ of Eq.~\protect\eqref{eq:func_g} is
 plotted as a function of $\tilde{a}$.
 (c) The boundary angle $\phi_0$ is plotted as a function of the
 thickness of the electrode $d$.
 (d) The analytical solution of the angle $\phi$
 in the top electrode given by Eqs.~\eqref{eq:phi_bessel} 
 and (\ref{eq:phi_zer_Bloch}) is 
 plotted by the solid line against the distance $r$ from the center of the
 contact. The result obtained by the Monte Carlo simulation is 
 plotted by the dotted line. The parameters are given in the main text. } 
\label{fig:fig2}
\end{figure}

We introduce the characteristic length\cite{MicromagneticBook} 
determined by the competition between the exchange interaction energy and 
the Zeeman energy as $\ell =\sqrt{2A/\mu_0 M_{\rm s}H}$ 
and use the superscript ``$\sim$'' to indicate the normalized values such as $\tilde{r}=r/\ell$,
$\tilde{a}=a/\ell$, $\tilde{d}=d/\ell$, and  $\tilde{h}=h/\ell$.
Then Eq.~\eqref{eq:euler1} is expressed as
\begin{equation}
\label{eq:euler2}
\ddot{\phi}
 +
 \frac{1}{\tilde{r}}\dot{\phi}
 -
\sin\phi
=0,
\end{equation}
where $\ddot{\phi}=\partial^{2} \phi / \partial {\tilde{r}}^{2}$ and $\dot{\phi}=\partial \phi / \partial {\tilde{r}}$.
Assuming that $\phi \lesssim \pi/4$ in the top electrode, 
the last term of Eq.~\eqref{eq:euler2} can be approximated 
as $\sin\phi\simeq\phi$ and we have
\begin{equation}
\label{eq:euler3}
\ddot{\phi}
 +
 \frac{1}{\tilde{r}}\dot{\phi}
 -
\phi
=0.
\end{equation}
Equation \eqref{eq:euler3} is the zeroth-order modified Bessel equation
whose solutions are the zeroth-order modified Bessel function of the
first kind $I_{0}({\tilde{r}})$ and the second kind $K_{0}({\tilde{r}})$. 
Since $I_{0}({\tilde{r}})$ diverges at the limit of $\tilde{r}\to\infty$, 
the angle $\phi({\tilde{r}})$ should be expressed as
\begin{equation}
\label{eq:phi_bessel}
 \phi({\tilde{r}}) 
= \phi_0\frac{K_{0}({\tilde{r}})}{K_{0}(\tilde{a})}.
\end{equation}
Substituting the approximation $\cos\phi\simeq 1-\frac{1}{2}\phi^{2}$
and Eq.~\eqref{eq:phi_bessel} into Eq.~\eqref{eq:et_bloch} we have
\begin{equation}
E_{\rm T} =2\pi d\, G(\tilde{a})\, \phi_0^{2},
\label{eq:et_bloch2}
\end{equation}
where 
\begin{align}
G(\tilde{a})
& =\frac{\tilde{a} K_{1}({\tilde{a}})}{K_{0}({\tilde{a}})}.
\label{eq:func_g}
\end{align}
Here, $K_{1}({\tilde{r}})$ is the first modified Bessel function of the second
kind.

\begin{figure}[t]
\includegraphics[width=\columnwidth]{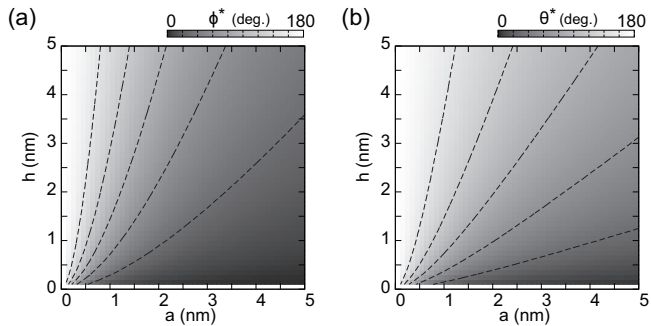}
\caption{(a) The twist angle of the Bloch wall is plotted as a function
 of the contact radius $a$ and contact height $h$. The interval of the
 contours is 30\degree.
 (b) The same plot of the N{\'e}el wall.} 
\label{fig:fig3}
\end{figure}

\begin{figure}[t]
\includegraphics[width=\columnwidth]{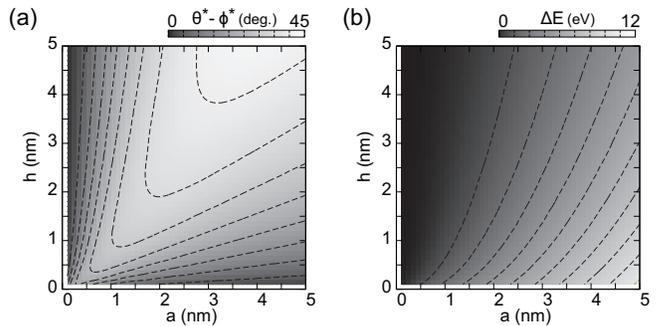}
\caption{(a) The difference between the twist angles $\theta^{\ast}$ and
 $\phi^{\ast}$ is
 plotted against the contact radius $a$ and the contact height $h$. 
 (b) The same plot for the energy difference between the N{\'e}el and the Bloch walls. }
\label{fig:fig4}
\end{figure}

In Fig. \ref{fig:fig2}~(a), we plot the energies $E_{\rm C}$
(dotted-line), $E_{\rm T}+E_{\rm B}$ (dashed-line) and $E_{\rm total}$
(solid-line) as functions of $\phi_0$. Here and hereafter 
we use the following parameters. 
The saturation magnetization is $M_{\rm s}=1.74\times 10^6$~A/m,  
the exchange stiffness constant $A=20.7$~pJ, the external field 
$H=7.96\times 10^4$~A/m ($1$~kOe), the thickness of the
electrodes $d=2$~nm, the height of the contact $h=2$~nm, and the radius
of the contact $a=1$~nm. The corresponding characteristic length is
$\ell = 15.4$ nm. Note that two parameters $a$ and $h$ will be varied in
Figs. \ref{fig:fig2}~(c), \ref{fig:fig3}~(a-b) and \ref{fig:fig4}~(a-b).
As shown in Fig. \ref{fig:fig2}~(a) $E_{\rm C}$
($E_{\rm T}+E_{\rm B}$) is a parabolic function of the boundary angle $\phi_0$
whose axis of symmetry is located at $\phi_0 = 0$ $(\pi/2)$, and 
$E_{\rm total}$ takes the minimum value at a certain value of $\phi_0$.
Setting the first derivative of $E_{\rm total}$ to be zero, we found that
the boundary angle of the ground state is
\begin{equation}
\phi_0 
 =
 \frac{\pi}{2}
 \left[
  \frac{{\tilde{a}}^{2}}{{\tilde{a}}^{2} + {\tilde{d}}\, {\tilde{h}}\, G(\tilde{a})}
\right].
\label{eq:phi_zer_Bloch}
\end{equation}
For the nano-contacts with $\tilde{a}\simeq\tilde{h}$, one can easily
see that the boundary angle $\phi_0$ depends strongly on the value of
$\tilde{d}$ because the function $G(\tilde{a})\simeq \tilde{a}$ as shown
in Fig. \ref{fig:fig2}~(b). 
In the limit of $\tilde{d}\to\infty$ the boundary angle $\phi_0$ converges to
zero, which means that the magnetic wall is perfectly confined in the
contact as expected from the discussions in Ref.~\onlinecite{bruno1999}. 
The value of $\phi_0$ increases with decreasing 
$\tilde{d}$ and reaches $\pi/2$ at the limit of $\tilde{d}\to 0$.
In Fig. \ref{fig:fig2}~(c) we plot the value of $\phi_0$ 
as a function of the thickness of the electrode $d$. 
If we assume that the thickness of the electrode is the same as the
height of the contact; i.e, $d=2$ nm, the boundary angle is as large as
$\phi_0=37.7\degree$. 

We also performed Monte Carlo\cite{MonteCarloBook} (MC) simulation to confirm the accuracy of the
analytical formula we derived.  
In the MC simulation, the width and depth of the electrode $L$ 
were set to be $L=100$~nm, which is large enough to eliminate the size dependence.
As shown in Fig. \ref{fig:fig2} (d) the angle $\phi(r)$ obtained by 
the MC simulation was well reproduced by our analytical formula of
Eq.~\eqref{eq:phi_bessel} with Eq.~(\ref{eq:phi_zer_Bloch}).

Let us move on to the N{\'e}el wall where we assume that the azimuthal angle 
$\phi = 0$, $d\phi/dz=0$, and $\dot{\phi}=0$. We also assume 
that $-\pi/2\le \theta \le \pi/2$. 
The energy of the contact is given by
\begin{equation}
\label{eq:ec_neel}
E_{\rm C}
 =\pi a^{2}\int_{-h/2}^{h/2} dz\, A
 \left(\frac{\partial \theta}{\partial z}\right)^{2}
 =\frac{2\pi a^{2} A}{h}(\pi-2\theta_0)^{2},
\end{equation}
where we have assumed that $\theta|_{z=h/2}=\pi/2-\theta_0$ 
and $\theta|_{z=-h/2}=\theta_0-\pi/2$. Note that Eq.~(\ref{eq:ec_neel}) 
is equivalent to Eq.~(\ref{eq:ec_bloch}). 
By rewriting $\theta$ in Eq.~(\ref{eq:energy_top_electrode}) 
with $\delta \theta\equiv \pi/2-\theta$ and taking up to 
the second order of $\delta \theta$, we obtain
\begin{align}
E_{\rm T}
 =2\pi d\int_{a}^{\infty} dr\, r
 \left[
  A \dot{\delta\theta}^{2} 
 + \frac{\mu_0M_{\rm s} H_{\rm eff}}{2}\delta\theta^{2}
\right], 
\label{eq:et_neel2}
\end{align} 
where $H_{\rm eff} \equiv H + M_{\rm s}$. 
The Euler equation for the N{\'e}el wall is given by
\begin{equation}
\label{eq:el_neel}
\ddot{\delta\theta}
 +
 \frac{1}{r}\dot{\delta\theta}
 -
\frac{\mu_0M_{\rm s} H_{\rm eff}}{2 A}\delta\theta
=0.
\end{equation}
The boundary conditions are $\delta\theta|_{r=a} = \theta_0$
and $\delta\theta|_{r\to\infty}=0$ in the top electrode, and
$\delta\theta|_{r=a} = \pi-\theta_0$ and
$\delta\theta|_{r\to\infty}=\pi$ in the bottom electrode.
Since Eq.~\eqref{eq:el_neel} takes the same form as Eq.~\eqref{eq:euler3}
except that $H$ is replaced with $H_{\rm eff}$, we can calculate $\delta \theta$ 
in a similar way as we calculate the Bloch wall, by 
replacing the characteristic length $\ell$ with 
$\ell_{\rm eff} \equiv \sqrt{2A/\mu_0M_{\rm s}H_{\rm eff}}$. Assuming the same parameters as in
Fig.~\ref{fig:fig2}~(a) $\ell_{\rm eff}$ is estimated to be 3.1 nm.
The energy of the top electrode $E_{\rm T}$ and 
the boundary angle $\theta_0$ of the ground state of the N{\'e}el wall 
are calculated from Eqs.~(\ref{eq:et_bloch2}) and (\ref{eq:phi_zer_Bloch}), 
respectively, by changing the length for the normalization 
from $\ell$ to $\ell_{\rm eff}$. 

In order to discuss the difference of the magnetic structures between the
Bloch wall and the N{\'e}el wall we introduce the twist
angles $\phi^{\ast}$ and $\theta^{\ast}$, which are defined as
$\phi^{\ast}\equiv\bigl|\phi|_{z=h/2} - \phi|_{z=-h/2}\bigr|=\pi-2\phi_0$ for the Bloch wall and 
$\theta^{\ast}\equiv\bigl|\theta|_{z=h/2} - \theta|_{z=-h/2}\bigr|=\pi-2\theta_0$ 
for the N{\'e}el wall. As shown in Figs. \ref{fig:fig3}~(a) and (b) $\theta^{\ast}$ is
always larger than $\phi^{\ast}$ because the penetration of the
N{\'e}el wall costs more demagnetization energy in the electrodes compared
with that of the Bloch wall. Since the resistance of the magnetic wall is
proportional to the twist angle\cite{levy1997}, the conditions for maximizing
the power of the current-induced oscillation between the Bloch wall and
the N{\'e}el wall are the same as those for the difference between
$\theta^{\ast}$ and $\phi^{\ast}$. In Fig. \ref{fig:fig4}~(a) we plot the
difference $\theta^{\ast} - \phi^{\ast}$ as a function of $a$ and
$h$. One can see that the optimum conditions for maximizing the 
power of the STO based on the nano-contact spin valve are approximately
given by $a \sim h$. 

Figure \ref{fig:fig4} (b) shows the difference between the energies
of the N{\'e}el wall and the Bloch wall defined as 
$\Delta E \equiv E_{\mbox{\scriptsize N{\'e}el}} - E_{\rm Bloch}$ as a function of $a$ and $h$. 
The energy difference $\Delta E$, and therefore the thermal stability
of the Bloch wall, increases with increasing $a$ or decreasing $h$.
In the plotted region ($0<a,h\le 5$nm) the Bloch wall is the ground state and the
N{\'e}el wall the excited state, contrary to the results of the perfectly
confined magnetic wall shown in Ref.~\onlinecite{coey2001}. The energy difference
$\Delta E$ for the perfectly confined magnetic wall originates from
the demagnetization energy in the contact that we neglected. 
Following Ref.~\onlinecite{coey2001} we can estimate $\Delta E \simeq $0.16 eV for
a perfectly confined magnetic wall with $a=$1 nm and $h=$2 nm.
On the other hand, for the the nano-contact spin valve with thin ferromagnetic
electrodes, the main contribution to $\Delta E$ is from the
demagnetization energy in the electrodes.  The estimated value of
$\Delta E$ is about 3.0 eV, which is about 19 times larger than that of
the perfectly confined magnetic wall.

In summary, we theoretically studied the effects on a magnetic structure of a
nano-contact spin valve with thin ferromagnetic electrodes.
We derived analytical formulas for the magnetic configurations of the
Bloch wall and the N{\'e}el wall, and the corresponding energies.
We showed that, compared with the Bloch wall, the penetration of the
N{\'e}el wall into the electrodes is suppressed by the increase of the
demagnetization energy. We found the optimum conditions for maximizing the 
power of the STO based on the nano-contact spin valve are given by $a \sim h$.
We also found that the thermal stability of the magnetic wall increases
as the nano-contact's radius increases or height decreases.

\begin{acknowledgments}
The authors would like to thank Dr. K. Miyake, Prof. M. Doi,
 Prof. M. Sahashi, and 
Prof. K. Sasaki for their valuable discussions and comments. 
This work is supported by MEXT KAKENHI Number 21740279
and JSPS KAKENHI Number 23226001.
\end{acknowledgments}


\begin{thebibliography}{9}
\bibitem{hayashi2008}
M.~Hayashi, L.~Thomas, R.~Moriya, C.~Rettner, and S.~S.~P.~Parkin, 
Science {\bf 320}, 209 (2008).

\bibitem{allwood2005}
D.~A.~Allwood, G.~Xiong, C.~C.~Faulkner, D.~Atkinson, D.~Petit, 
and R.~P.~Cowburn, Science {\bf 309}, 1688 (2005).

\bibitem{garcia1999}
M.~M.~N.~Garc{\'i}a and Y.-W.~Zhao, Phys. Rev. Lett. {\bf 82}, 2923 (1999).

\bibitem{imamura2000}
H.~Imamura, N.~Kobayashi, S.~Takahashi, and S.~Maekawa, 
Phys. Rev. Lett. {\bf 84}, 1003 (2000).

\bibitem{takagishi2009}
M.~Takagishi, H.~N.~Fuke, S.~Hashimoto, H.~Iwasaki, S.~Kawasaki, 
R.~Shiozaki, and M.~Sahashi, J. Appl. Phys. {\bf 105}, 07B725 (2009).

\bibitem{He2007}
J. He and S. Zhang, Appl. Phys. Lett. {\bf 90}, 142508 (2007).

\bibitem{Ono2008}
T. Ono and Y. Nakatani, Appl. Phys. Express {\bf 1} 061301 (2008).

\bibitem{franchin2008}
M.~Franchin, T.~Fischbacher, G.~Bordignon, P.~de Groot,and H.~Fangohr, 
Phys. Rev. B {\bf 78}, 054447 (2008). 

\bibitem{matsushita09B}
K.~Matsushita, J.~Sato and H.~Imamura, J. Phys. Soc. Jpn. {\bf 78}, 093801 (2009).

\bibitem{Bisig2009}
A. Bisig, L. Heyne, O. Boulle, and M. Klaui, Appl. Phys. Lett. {\bf 95}, 162504 (2009). 

\bibitem{suzuki2009}
H.~Suzuki, H.~Endo, T.~Nakamura, T.~Tanaka, M.~Doi, S.~Hashimoto, 
H.~N.~Fuke, M.~Takagishi, H.~Iwasaki, and M.~Sahashi, 
J. Appl. Phys. {\bf 105}, 07D124 (2009).



\bibitem{ono1999} 
T.~Ono, Y.~Ooka, H.~Miyajima, and Y.~Otani, Appl. Phys. Lett. {\bf 75}, 1622 (1999).

\bibitem{bruno1999}
P.~Bruno, Phys. Rev. Lett. {\bf 83}, 2425 (1999).

\bibitem{Gorkom:1999}
 R.~P.~van~Gorkom, J.~Caro, S.~J.~C.~H.~Theeuwen, K.~P.~Wellock, 
N.~N.~Gribov, and S.~Radelaar, Appl. Phys. Lett. {\bf 74}, 422 (1999). 
 
\bibitem{coey2001}
J.~M.~D.~Coey, L.~Berger, and Y.~Labaye, 
Phys. Rev. B {\bf 64}, 020407(R) (2001).

\bibitem{chopra2005}
H.~D.~Chopra, M.~R.~Sullivan, J.~N.~Armstrong, and S.~Z.~Hua, 
Nature Materials {\bf 4}, 832 (2005).



\bibitem{matsushita09A}
K.~Matsushita, J.~Sato and H.~Imamura, J. Appl. Phys. {\bf 105}, 07D525 (2009).

\bibitem{matsushita10}
K.~Matsushita, J.~Sato, H.~Imamura, and M.~Sasaki, J. Phys. Soc. Jpn. {\bf 79}, 093801 (2010). 

\bibitem{arai2012}
H.~Arai, H.~Tsukahara, and H.~Imamura, Appl. Phys. Lett. {\bf 101}, 092405 (2012).



\bibitem{Molyneux02}
V.~A.~Molyneux, V.~V.~Osipov, and E.~V.~Ponizovskaya, Phys. Rev. B {\bf 65}, 184425 (2002). 

\bibitem{Jubert05}
P.-O.~Jubert and R. Allenspach, J. Magn. Magn. Mater. {\bf 290-291}, 758 (2005). 

\bibitem{KohnSlastikov06}
R.~V.~Kohn and V.~V.~Slastikov, Calc. Var. Part. Diff. Equ. {\bf 28}, 33 (2006).

\bibitem{Fuke07}
H.~N.~Fuke, S.~Hashimoto, M.~Takagishi, H.~Iwasaki, S.~Kawasaki, 
K.~Miyake, and M.~Sahashi, IEEE Trans. Magn. {\bf 43}, 2848 (2007). 

\bibitem{takagishi2010}
M.~Takagishi, K.~Yamada, H.~Iwasaki, H.~N.~Fuke, and S.~Hashimoto, 
IEEE Trans. Magn. {\bf 46}, 2086 (2010).

\bibitem{MicromagneticBook} H. Kronm{\"u}ller and M. F{\"a}hnle, 
{\it Micromagnetism and the Microstructure of Ferromagnetic Solids} 
(Cambridge University Press, Cambridge, 2003) Chap.~13.


\bibitem{MonteCarloBook} D.~P.~Landau and K.~Binder, 
{\it A Guide to Monte Carlo Simulations in Statistical Physics (3rd ed.)} 
(Cambridge University Press, Cambridge, 2009).


\bibitem{levy1997}
P.~M.~Levy and S.~Zhang, Phys. Rev. Lett. {\bf 79}, 5110 (1997).

\end{thebibliography}
\end{document}